\documentclass[twocolumn,amsmath,amssymb,prb,floatfix,superscriptaddress]{revtex4-2}

\usepackage{graphicx}
\usepackage{dcolumn}
\usepackage{bm}
\usepackage{hyperref}
\usepackage{xcolor}
\usepackage{times}
\usepackage{float}
\usepackage{placeins}
\hypersetup{ 
    colorlinks,
    citecolor=blue,
    filecolor=blue,
    linkcolor=blue,
    urlcolor=blue 
}

\begin{document}

\title{Symmetric projected entangled-pair states analysis of a phase transition in coupled spin-1/2 ladders}

\author{Juraj Hasik}
\affiliation{Institute for Theoretical Physics, University of Amsterdam, Science Park 904, 1098 XH Amsterdam, The Netherlands}
\affiliation{Laboratoire de Physique Th\'eorique, Universit\'e de Toulouse, CNRS, UPS, France}
\author{Glen Bigan Mbeng}
\affiliation{Institut f\"ur Theoretische Physik, Universit\"at Innsbruck, A-6020 Innsbruck, Austria}
\author{Sylvain Capponi}
\affiliation{Laboratoire de Physique Th\'eorique, Universit\'e de Toulouse, CNRS, UPS, France}
\author{Federico Becca}
\affiliation{Dipartimento di Fisica, Universit\`a di Trieste, Strada Costiera 11, I-34151 Trieste, Italy}
\author{Andreas M. L\"auchli}
\affiliation{Laboratory for Theoretical and Computational Physics, Paul Scherrer Institute, 5232 Villigen, Switzerland}
\affiliation{Institute of Physics, \'{E}cole Polytechnique F\'{e}d\'{e}rale de Lausanne (EPFL), 1015 Lausanne, Switzerland}
\affiliation{Institut f\"ur Theoretische Physik, Universit\"at Innsbruck, A-6020 Innsbruck, Austria}

\date{\today}

\begin{abstract}
Infinite projected entangled-pair states (iPEPS) have been introduced to accurately describe many-body wave functions on two-dimensional lattices. In this context, 
two aspects are crucial: the systematic improvement of the {\it Ansatz} by the optimization of its building blocks, i.e., tensors characterized by bond dimension 
$D$, and the extrapolation scheme to reach the ``thermodynamic'' limit $D \to \infty$. Recent advances in variational optimization and scaling based on correlation 
lengths demonstrated the ability of iPEPS to capture the spontaneous breaking of a continuous symmetry in phases such as the antiferromagnetic (N\'eel) phase with 
high fidelity, in addition to valence-bond solids which are already well described by finite-$D$ iPEPS. In contrast, systems in the vicinity of continuous quantum 
phase transitions still present a challenge for iPEPS, especially when non-abelian symmetries are involved. Here, we consider the iPEPS {\it Ansatz} to describe 
the continuous transition between the (gapless) antiferromagnet and the (gapped) paramagnet that exists in the $S=1/2$ Heisenberg model on coupled two-leg ladders. 
In particular, we show how accurate iPEPS results can be obtained down to a narrow interval around criticality and analyze the scaling of the order parameter in 
the N\'eel phase in a spatially anisotropic situation. 
\end{abstract}

\maketitle

\section{Introduction}\label{sec:intro}

One of the most challenging problem in condensed matter theory is to obtain sufficiently accurate approximations of the ground state and low-energy excitations of 
generic many-body Hamiltonians. For this reason, several numerical approaches have been devised in the last 30 years, including density-matrix renormalization group 
(DMRG)~\cite{white1992}, dynamical mean-field theory~\cite{georges1992}, and quantum Monte Carlo (QMC) techniques~\cite{white1989}. Among them, the natural extensions 
of DMRG, such as matrix-product states~\cite{ostlund1995,perez2007} in one dimension and projected entangled-pair states (PEPS)~\cite{verstraete2004a,verstraete2004b} 
in two dimensions, represent a promising computational framework to get excellent approximations of strongly-correlated lattice models. Within this approach, the 
ground-state wave function is represented by means of local tensors, typically associated to the sites of the underlying lattice. These tensors have two kinds of 
indices: a single {\it physical} index specifying the local physical configuration (e.g., $S^z=\pm 1/2$ in a spin-$1/2$ model) and a collection of {\it bond} indices 
(whose number, usually, equals the the coordination number of the lattice), each one having dimension $D$. This latter quantity controls the accuracy of the 
{\it Ansatz}, the exact ground-state wave function being achieved for $D \to \infty$. The bond indices on nearest-neighbor sites are contracted together, thus 
defining a {\it tensor network}. Furthermore, infinite systems may be considered by embedding the tensor network within a suitable environment. Such tensor networks,
dubbed iPEPS, are the main focus of this work. 

In the last decade, several algorithms to optimize iPEPS {\it Ans\"atze} have been developed and used to address challenging problems in different correlated systems, 
ranging from the characterization of unconventional states of matter (e.g., spin liquids in highly-frustrated spin model) to the competition between stripes and 
superconductivity in Hubbard or $t{-}J$ models. For example, the ground state of the spin-$1/2$ Heisenberg model has been analysed on the kagome~\cite{liao2017} 
and Shastry-Sutherland~\cite{corboz2013} lattices. Hamiltonians with an enlarged $SU(N)$ ``spin'' symmetry have been also considered, to assess both magnetically 
ordered and disordered phases~\cite{corboz2011,corboz2012}. In this regard, a particular attention has been devoted to the possibility to stabilize chiral spin 
liquids~\cite{chen2020,chen2021}. As far as electronic models are concerned, the evidence in favor of stripes has been pushed for $t{-}J$~\cite{corboz2014} and
Hubbard~\cite{zheng2017} models, in the vicinity of the hole doping $\delta=1/8$.

By construction, the iPEPS {\it Ansatz} has been devised to describe both gapped and gapless phases, satisfying the area law~\cite{eisert2010}; however, within the 
optimization procedure in generic systems, gapless states are not easily obtained. Similarly, it is also not easy to correctly describe states in the vicinity of 
continuous quantum phase transitions. This is in a stark contrast with what happens in one spatial dimension, where tensor-network states, equipped with an appropriate 
scaling theory, may capture very well gapless phases and critical phenomena~\cite{pollmann2009,pirvu2012,rams2018,vanhecke2019}. Recently, a progress in the iPEPS 
analysis of critical and/or gapless systems has been made thanks to two key developments. The first one is the introduction of optimization techniques, based either 
on diagram summations~\cite{corboz2016,vanderstraeten2016} or the so-called automatic differentiation~\cite{liao2019}, which substantially improve the accuracy with 
respect to the commonly used imaginary time evolution methods. The second one is the development of finite-correlation-length scaling (FCLS)~\cite{rader2018,corboz2018} 
that allowed to leverage the well-established finite-size scaling approach to iPEPS states. In fact, the accuracy of thermodynamic estimates based on finite-$D$ 
iPEPS calculations can be considerably improved. The efficiency of these advances was recently demonstrated on a paradigmatic problem of the spin-$1/2$ Heisenberg 
$J_1-J_2$ model on a square lattice by estimating the magnetization curve within the N\'eel phase, even in the vicinity of the transition to the quantum spin-liquid 
state~\cite{hasik2021}.

In this work, we pursue the idea of describing a continuous phase transition within the iPEPS formalism. In particular, as for the $J_1-J_2$ model, we focus on a 
case where the N\'eel order is melted by quantum fluctuations, namely a two-dimensional system of coupled spin-$1/2$ Heisenberg ladders defined by
\begin{equation}\label{eq:ham}
{\cal H} = J \sum_{R} {\bf S}_{R} \cdot {\bf S}_{R+\hat{x}} + \sum_{R} J_{R} {\bf S}_{R} \cdot {\bf S}_{R+\hat{y}},
\end{equation}
where ${\bf S}_{R}=(S^x_{R},S^y_{R},S^z_{R})$ is the $S=1/2$ operator on the site $R=(x,y)$ of a square lattice, $\hat{x}$ and ${\hat{y}}$ are unit vectors in $x$ 
and $y$ direction, and $J_{R}=J$ or $J_{R}=\alpha J$, depending on the parity of $y$. By varying $\alpha$, this model interpolates between the Heisenberg model on 
the square lattice at $\alpha=1$ and a system of decoupled two-leg ladders at $\alpha=0$. In the following, we will take $J$ as the energy scale. Owing to the absence 
of the sign problem, this system can be studied by unbiased QMC techniques and, therefore, offers an excellent benchmark for the accuracy of iPEPS to describe a 
non-trivial quantum phase transition, e.g., beyond the simplest case of the Ising model in transverse field. Indeed, the Hamiltonian~\eqref{eq:ham} displays a quantum 
critical point at $\alpha_c=0.31407(5)$, separating a gapless antiferromagnet and a gapped paramagnet~\cite{matsumoto2001}. The critical exponents are compatible with 
the ones of the classical three-dimensional $O(3)$ Heisenberg model, as expected.

A recent iPEPS investigation of this Heisenberg model~\cite{hasik2019} highlighted the difficulties faced by the unrestricted optimizations of iPEPS tensors across 
the phase transition. In fact, due to the unbroken $SU(2)$ symmetry in the paramagnetic side, strong finite-$D$ effects are present, thus impeding a systematic 
analysis. Moreover, the optimization procedure is problematic also within the antiferromagnetic phase, where the expected $U(1)$ symmetry around the direction of 
the staggered magnetization is usually broken. In this work, we want to constrain the iPEPS {\it Ansatz} by imposing this $U(1)$ symmetry, thus limiting finite-$D$
effect, and optimize such symmetric tensors with the automatic differentiation~\cite{liao2019}. In addition, we extend the FCLS analysis~\cite{rader2018,corboz2018} 
in situations with a spatial anisotropy, due to the presence of two length scales, since the correlations along the spatial $x$ and $y$ directions will be generically 
different in the ground state of Eq.~\eqref{eq:ham}. These two improvements allow us to get accurate results for the antiferromagnetic order parameter up to the 
critical point, as confirmed by a direct comparison with QMC calculations. Finally, we included an external staggered magnetic field in the Hamiltonian~\eqref{eq:ham}, 
which allows us to inspect the response in the order parameter, even very close to the critical point.

\section{Methods}\label{sec:methods}

Here, we briefly describe the structure of the iPEPS {\it Ansatz} that is used to investigate the Hamiltonian~\eqref{eq:ham}. In order to account for both the 
antiferromagnetic order in the gapless regime and the short-range valence-bond correlations in the gapped one, we consider four rank-5 tensors $t=\{a,b,c,d\}$, 
each one with a physical index $s$ (with dimension $2$, suitable for $S=1/2$ degrees of freedom) and four auxilliary indices $u,l,d,r$ (with dimension $D$). These 
tensors are arranged in a $2\times2$ unit cell to tile the square lattice:
\begin{equation}\label{eq:ansatz}
|\textrm{iPEPS}(a,b,c,d)\rangle = \vcenter{\hbox{\includegraphics[scale=0.25]{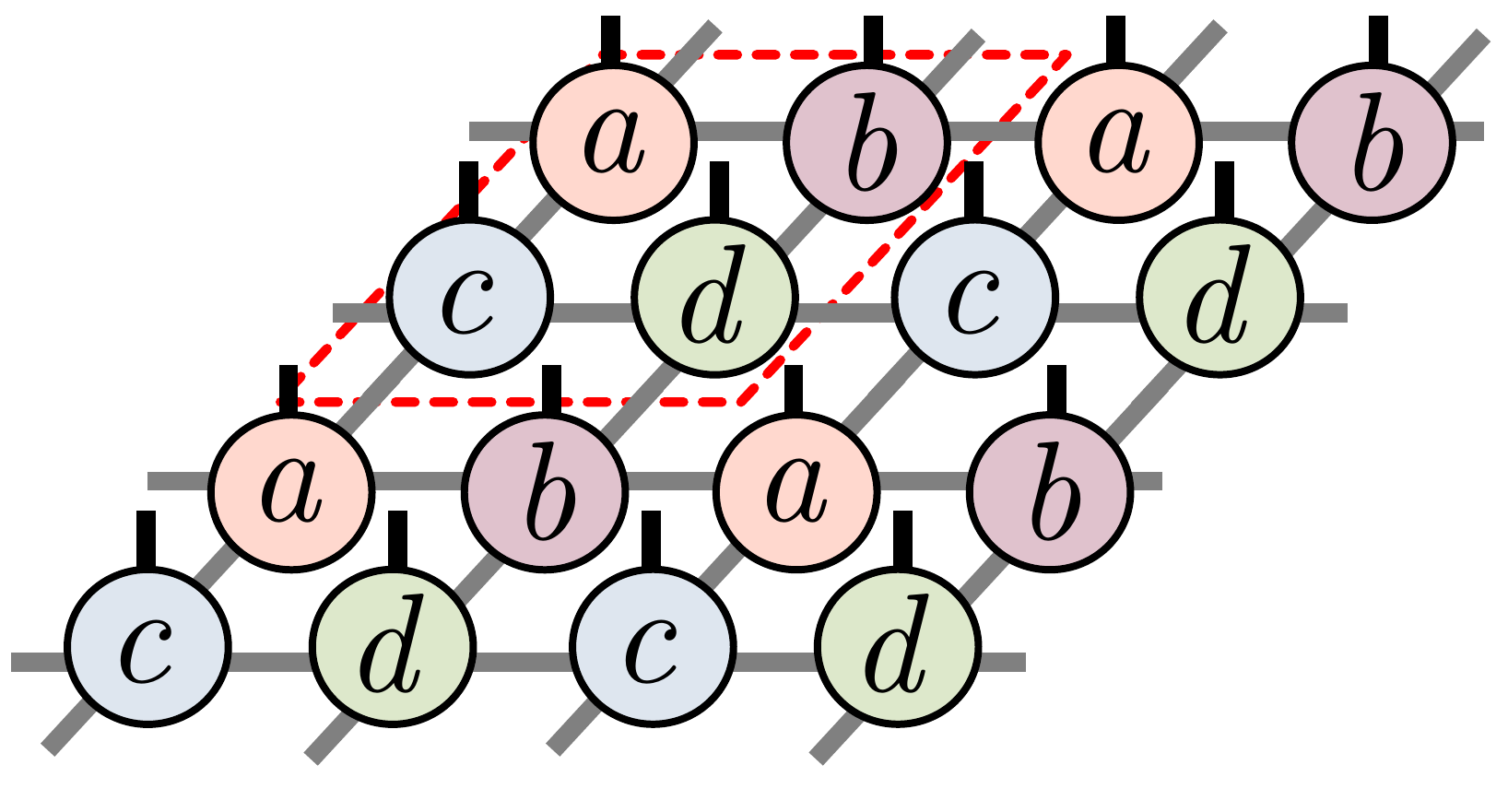}}}.
\end{equation}
Observables are obtained by computing effective environments within the corner-transfer matrix method, generalized for extended unit cells~\cite{corboz2014}.

The gapped phase retains all the symmetries of the Hamiltonian and, in particular it has the full $SU(2)$ spin symmetry; by contrast, the gapless phase breaks the 
spin-rotational symmetry, leading to a finite collinear magnetization and a residual $U(1)$ symmetry about it. Therefore, we impose an explicit $U(1)$ spin symmetry 
on the tensors. This can be achieved by associating integer ``charges''  to the two physical spin-$1/2$ components and the $D$ virtual degrees of freedom (on each 
of the bond indices)~\cite{hasik2019}. To preserve point-group symmetries of the underlying lattice, a straightforward choice is to take the same charges on all 
the bond indices that are equivalent under the symmetries (e.g., the $C_{4v}$ symmetry used in Ref.~\cite{hasik2019} led to such choice). Here, we do not want to 
impose any symmetry within the $2\times 2$ unit cell, which allows the possibility that different charges are defined on different indices. Thus, each of the four 
tensors will possess a set of charges ${\vec v}^{\gamma}_j=(v_0,\dots,v_{D-1})$ with $\gamma=a,b,c,d$ and $j=u,l,d,r$. Instead, the same charges 
${\vec u}=(u^{\uparrow},u^{\downarrow})$ for the physical indices are taken for all tensors. Therefore, the $U(1)$ symmetry is realized by enforcing a selection 
rule for the non-zero elements of the tensors $t^s_{uldr}$:
\begin{equation}
u^{s} + v^{\gamma}_u + v^{\gamma}_l + v^{\gamma}_d + v^{\gamma}_r = N,
\end{equation}
where, without loss of generality, we chose the case with $N=0$ to work with invariant tensors.

The definition of these charges allows us to improve the numerical efficiency, since tensors may arranged in blocks with definite values of them, allowing a 
block-sparse representation. For a detailed treatment of tensor networks with $U(1)$ symmetry and linear algebra with block-sparse tensors, see Ref.~\cite{singh2011}. 
Importantly, once the blocks are defined, the gradient optimization changes the tensors elements without mixing different blocks. The implementation of linear 
tensor algebra for abelian-symmetric tensors is provided by open-source library {\it YAST}~\cite{yast} and the iPEPS algorithms built on top of it are available 
in the {\it peps-torch} library~\cite{pepstorch}.

From a practical point of view, one possibility to identifying $U(1)$ charges (for the best variational states) is to perform imaginary-time evolution at different 
values of $\alpha$ by using a two-site simple-update scheme~\cite{jiang2008} starting from product states with $U(1)$ symmetry. Remarkably, this approach does not 
lead to substantial improvements (in terms of physical quantities, as energy or magnetization) with respect to the charges obtained from the unrestricted optimizations 
of single-site iPEPS {\it Ansatz} with the $C_{4v}$ lattice symmetry in the Heisenberg model with $\alpha=1$~\cite{hasik2021}, see Appendix~\ref{sec:app-u1classes}. 

\section{Results}\label{sec:results}

Let us now show the results obtained within the optimization of the iPEPS {\it Ans\"atze}. In order to perform a careful comparison with QMC calculations (which are 
numerically exact, since the spin model does not suffer from the sign problem), we apply the stochastic series expansion method~\cite{sandvik1999} and perform 
extrapolations by decreasing the temperature (to reach ground-state properties) and increasing the size of the cluster (to reach the thermodynamic limit). Most of 
the QMC calculations were done on $L\times L$ lattices with periodic boundary conditions, at temperature $T=1/(2L)$, by using the ALPS libraries~\cite{ALPS13,ALPS2}. 

\subsection{Phase diagram}

\begin{figure}
\includegraphics[width=\columnwidth]{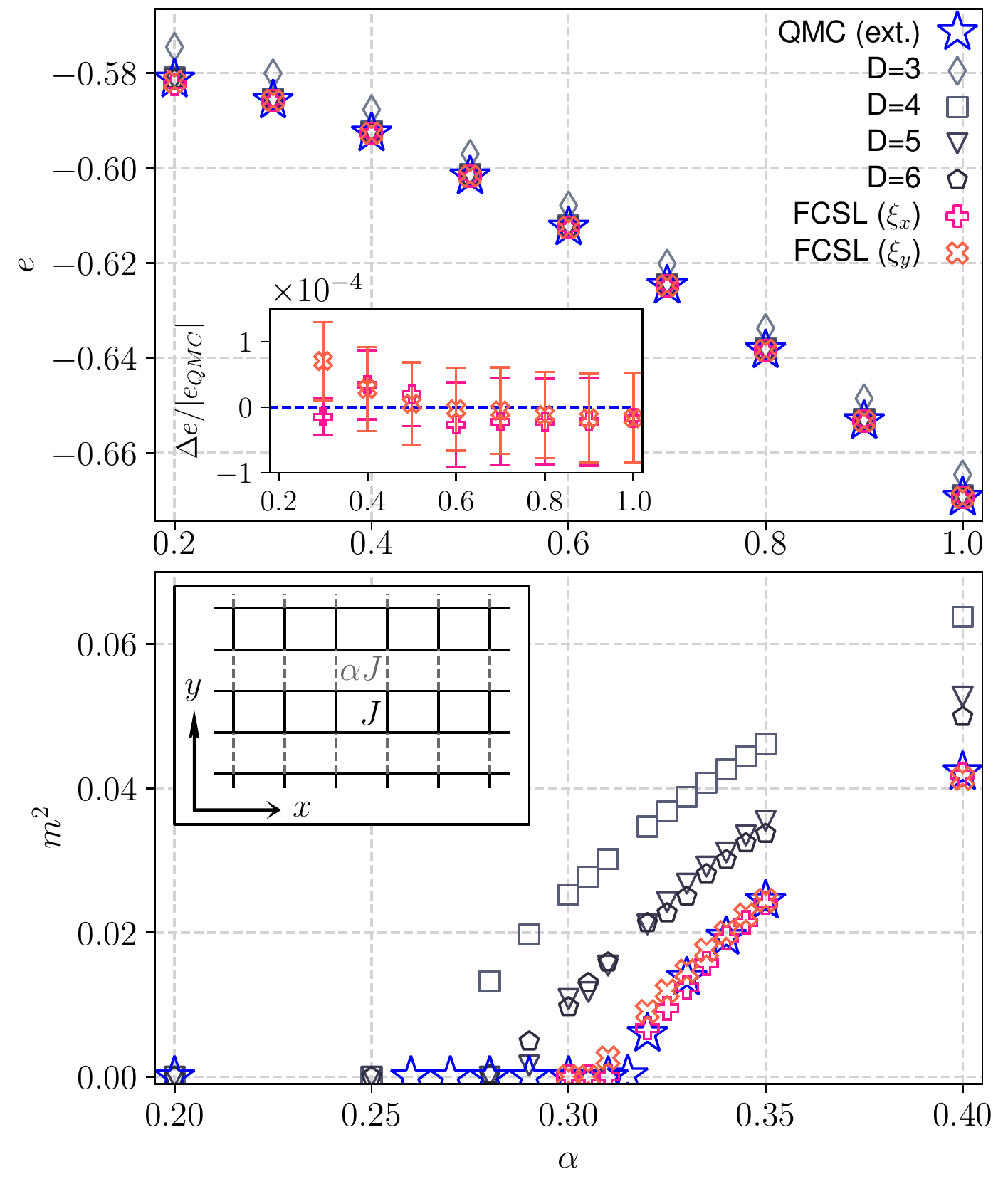}
\caption{\label{fig:phasediag} 
Energy per site (upper panel) and staggered magnetization (lower panel) for optimized iPEPS with $U(1)$ symmetry and $2\times2$ unit cell. The results for $D=3,\ldots,6$
are reported, as well as the ones obained with the FCLS extrapolation, see text. In addition, QMC results are also shown for the thermodynamic extrapolation. The inset 
in the upper panel shows the relative error of the iPEPS energies (with FCLS extrapolations) with respect to the QMC ones (with thermodynamic extrapolations). The inset
in the bottom panel gives a sketch of the super-exchanges in Hamiltonian~\eqref{eq:ham}, where intra-ladder $J$ (inter-ladder $\alpha J$) couplings are denoted by 
black solid (gray dashed) lines.}
\end{figure}

In Fig.~\ref{fig:phasediag}, we report the energy per site $e$ and the staggered magnetization $m^2$ (averaged over the $2\times 2$ unit cell), for different 
values of $\alpha$ across the quantum critical point. The outcome shows that, for small bond dimension, i.e., $D \le 3$, the accuracy is quite poor in both the 
antiferromagnetic and paramagnetic phases. However, once the bond dimension becomes large enough to capture the entanglement structure of the ground state, the 
tensor network provides an excellent variational description, not only in the paramagnetic phase, where the iPEPS parametrization is particularly suitable, but
also within the magnetically ordered phase. The accuracy of the ground-state energy is exemplified by showing the relative error with respect to the value $e_{QMC}$, 
obtained within the QMC approach (extrapolated in the thermodynamic limit). Indeed, the quantity $\Delta e/|e_{QMC}|$, with $\Delta e=|e-e_{QMC}|$ is vanishing 
within the statistical errors, see Fig.~\ref{fig:phasediag}. Most importantly, the magnetization curves for finite values of the bond dimension $D$ follow the QMC 
data faithfully in the bulk of each phase, still overestimating the order parameter close to the critical point. We would like to mention that the present QMC 
results allow us to locate the critical point at $\alpha_c=0.31467(1)$, which is slightly different from the previous estimate $\alpha_c=0.31407(5)$~\cite{matsumoto2001}
(see Appendix~\ref{sec:app-qmc-alphac}).

\subsection{Finite-correlation-length scaling}

Obtaining accurate and reliable estimates of the observables (e.g., the energy per site or the staggered magnetization) within iPEPS {\it Ans\"atze} represents an 
important issue that has been addressed since their definition. Recently, an interesting step forward has been obtained for magnetic states that break a continuous 
symmetry (e.g., the antiferromagnetic one in the Heisenberg model). In this case, the presence of gapless modes induces a diverging correlation length. When working 
on finite clusters, as in standard QMC approaches, this fact translates into well-defined size-scaling laws of the physical quantities~\cite{neuberger1989,fisher1989}. 
Instead, within iPEPS, we directly work in the thermodynamic limit (that is mimicked by the embedding procedure) and, therefore, it is not possible to straightforwardly 
apply the same scaling laws; still, the correlation length $\xi$ is finite for any finite value of the bond dimension $D$ and diverges with $D\to \infty$. As a result, 
it has been suggested~\cite{rader2018,corboz2018} that the FCLS analysis can be defined in terms of $\xi$ instead of the cluster size as
\begin{eqnarray}
\label{eq:fclse}
e(\xi)   &=& e_0 + \frac{a}{\xi^3} + O\left(\frac{1}{\xi^4}\right), \label{eqn:FCLS_erg}\\
m^2(\xi) &=& m^2_0 + \frac{b}{\xi} + O\left(\frac{1}{\xi^2}\right), \label{eq:fclsm}
\end{eqnarray}
where $a$ and $b$ are suitable constants. The system we are interested in, i.e., the Heisenberg model on coupled two-leg ladders, presents a complication, as it 
breaks the $\pi/2$-rotational symmetry for $\alpha \neq 1$, leading to two distinct length scales. Within finite-$D$ iPEPS calculations, we have access to both of 
them by looking at the spectrum of the transfer matrices along $x$ (i.e., within ladders) and $y$ (i.e., across ladders). Thus, we obtain two correlation lengths 
($\xi_x$ and $\xi_y$) and perform FCLS independently for both of them. The extrapolated values $e_0$ and $m^2$ are independent of choice of the correlation length 
direction, while the parameters $a$ and $b$ depend on the choice. It must be emphasized that these FCLS relations are expected to be valid exclusively for a Goldstone
phase with a spontaneously broken continuous symmetry, where the correlation length diverges with increasing $D$. Instead, within the paramagnetic gapped phase, 
$\xi$ remains finite in the thermodynamic limit, thus implying that Eqs.~\eqref{eq:fclse} and~\eqref{eq:fclsm} cannot be applied. The impossibility to fit the 
numerical data for a given $\alpha$ within this scheme is then taken as the evidence that the state is not gapless (antiferromagnetic). At the quantum critical point,
we still expect Eq.~\eqref{eqn:FCLS_erg} to hold, while the order parameter scaling is replaced with a quantum critical form $m(\xi) \approx \xi^{-\Delta_m}$, with 
$\Delta_m$ the scaling dimension of the order parameter at the critical point~\cite{rader2018}.

\begin{figure}
\includegraphics[width=\columnwidth]{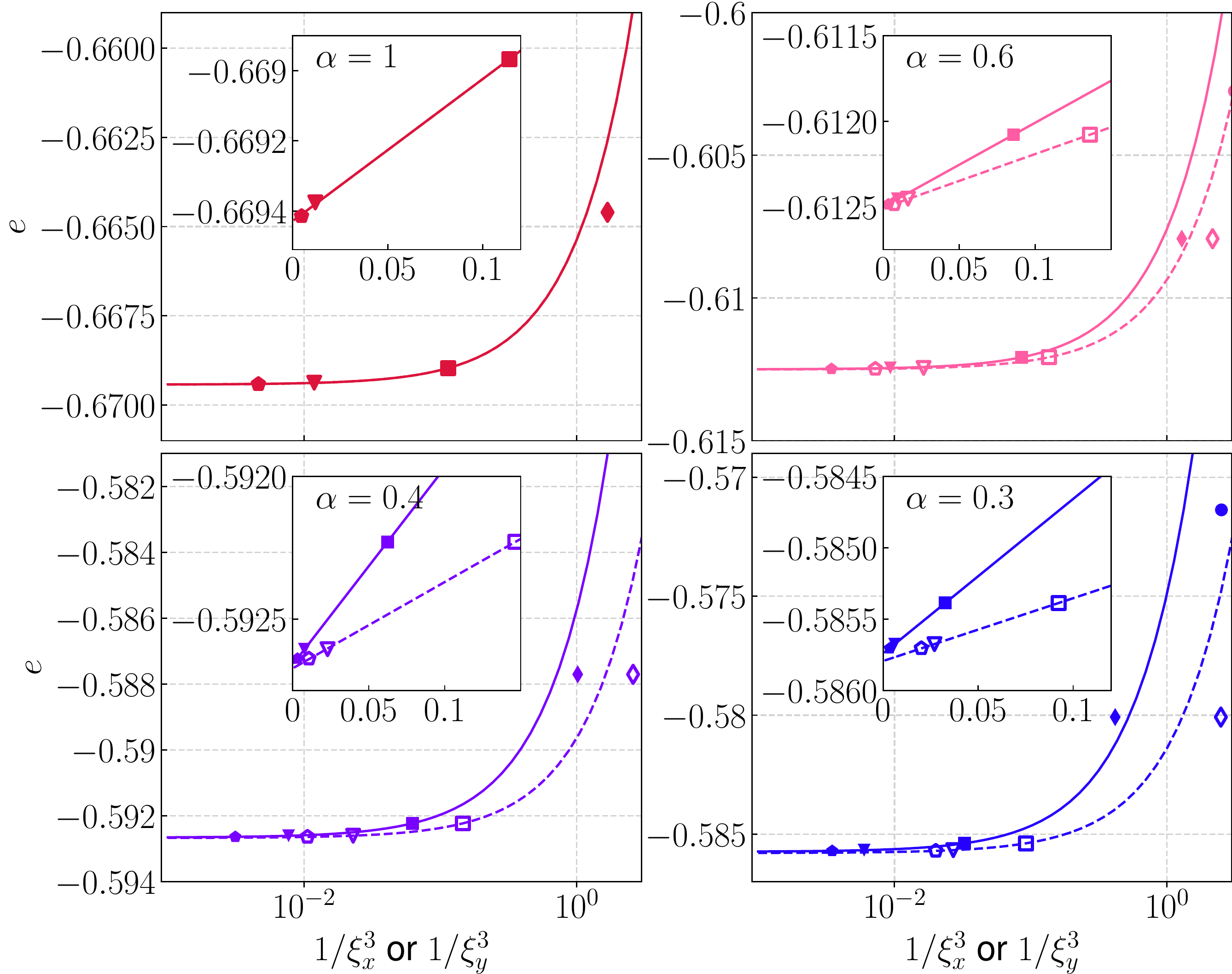}
\caption{\label{fig:fcls-e} 
FCLS extrapolations for the energy $e$, by using both horizontal ($\xi_x$, full symbols and lines) and vertical ($\xi_y$, empty symbols and dashed lines) correlation 
lengths, for a few values of the inter-ladder couplings $\alpha$. The symbols correspond to iPEPS with $D=3$ (diamonds), $D=4$ (squares), $D=5$ (triangles), and $D=6$ 
(pentagons). The colors follow the $\alpha$ color scheme of Fig.~\ref{fig:fcls-m2}.}
\end{figure}

\begin{figure}
\includegraphics[width=\columnwidth]{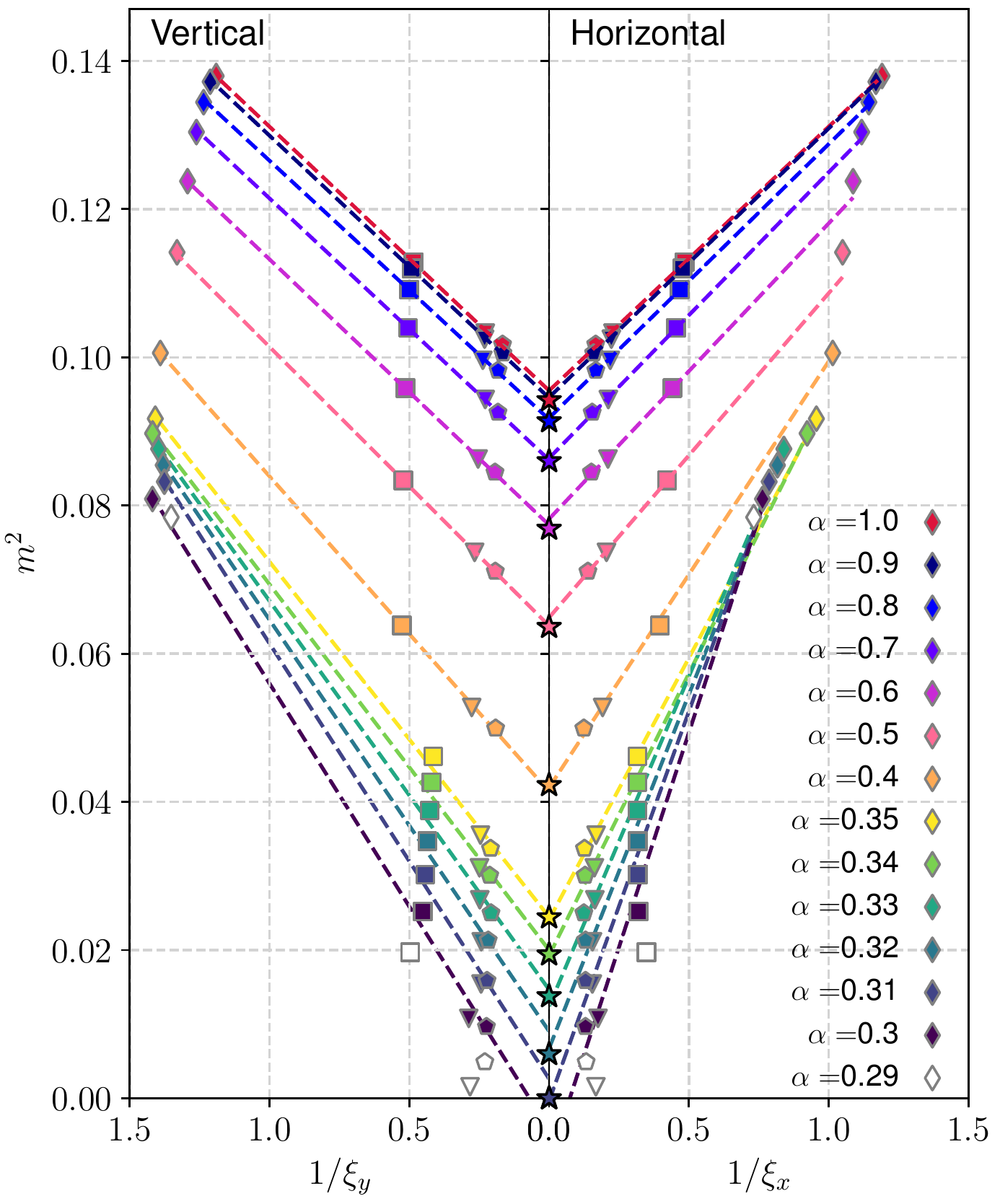}
\caption{\label{fig:fcls-m2} 
FCLS extrapolations for staggered magnetization $m^2$, by using both vertical ($\xi_y$, left part) and horizontal ($\xi_x$, right part) correlation lengths for 
$0.29 \le \alpha \le 1$. The symbols correspond to iPEPS with  $D=3$ (diamonds), $D=4$ (squares), $D=5$ (triangles), and $D=6$ (pentagons); thermodynamic extrapolations 
from QMC data are also reported (stars).}
\end{figure}

The results for the energy per site $e$ are reported for selected values of $\alpha$ in Fig.~\ref{fig:fcls-e}. Within the magnetically ordered phase (but also close 
to the quantum critical point), the extrapolations of the FCLS analysis using either $\xi_x$ or $\xi_y$ give consistent values, providing an internal verification of 
the approach. Most importantly, the staggered magnetization $m^2$ can be also extracted for infinite correlation length, see Fig.~\ref{fig:fcls-m2}. The extrapolated 
results are in very good agreement with QMC estimates, up to values of the inter-ladder couplings that are very close to the critical point. Beyond the critical point, 
the system develops a gap and thus the FCLS fails to provide the correct scaling, where correlation lengths remain finite. Nevertheless, within the paramagnetic regime, 
finite-$D$ iPEPS calculations are able to recover accurate results. Indeed, for $\alpha \lesssim 0.28$, a vanishing magnetization is already obtained with $D=5$. 
However, upon approaching criticality, i.e., at $\alpha=0.3$, finite-$D$ iPEPS retain a small residual magnetization (at least up to $D=6$). In fact, close to the 
critical point, the correlation length far surpasses the values of $\xi_x$ and $\xi_y$ that our finite-$D$ iPEPS can support. Hence, the FCLS analysis can be used to 
improve the estimates of the energy. Still, some important deviations in the scaling laws are clearly visible, see Fig.~\ref{fig:fcls-m2}. Finally, let us remark that 
the iPEPS optimization in the narrow region near criticality is technically difficult due to presence of instabilities (see Appendix~\ref{sec:opt-instability}).

\begin{figure*}
\includegraphics[width=\textwidth]{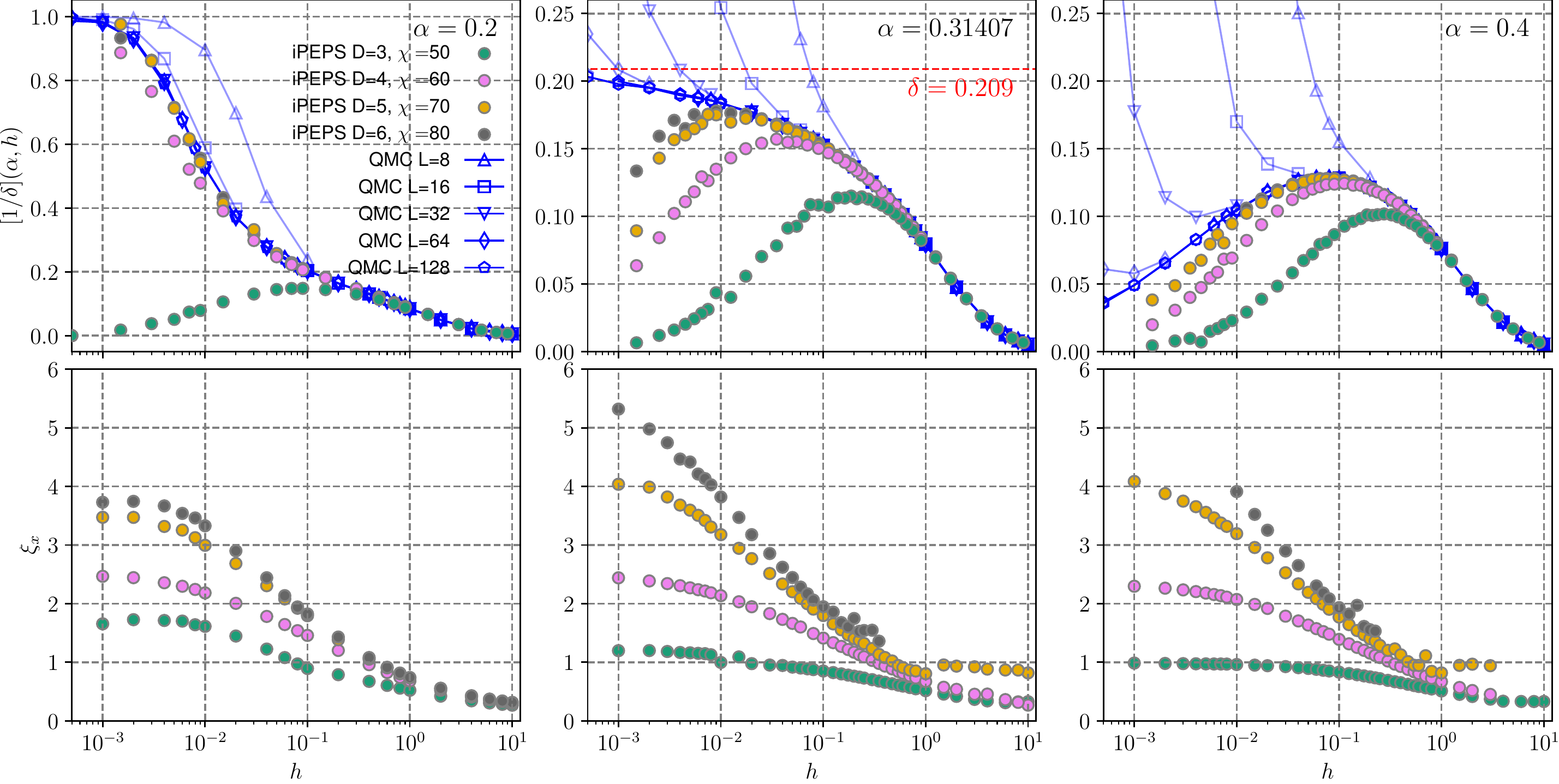}
\caption{\label{fig:hz-alpha} 
$U(1)$-symmetric iPEPS with $D=3,\ldots,6$ in the presence of external staggered field $h$. Upper panels: Running exponent as a function of $h$ for the gapped phase 
with $\alpha=0.2$ (left), at criticality $\alpha=0.31407$ (middle), and in the N\'eel phase $\alpha=0.4$ (right). The value of the critical exponent $1/\delta$ is 
reported at the critical point. QMC results are also reported for comparison, with faded colors marking the data affected by finite lattice size. Bottom panels: 
horizontal correlation length $\xi_x$ as a function of $h$.}
\end{figure*}

\subsection{The effect of a staggered magnetic field}

In the previous section, we focused on estimating observables of the ground state of the $SU(2)$-symmetric Hamiltonian~\eqref{eq:ham}. Here, we explicitly break this 
symmetry to investigate the phase transition with a complementary non-symmetric approach~\cite{binder1984,demidio2021,rader2018}. In particular, we follow the approach 
of Refs.~\cite{rader2018,demidio2021}, which recently used tensor networks to compute running exponents in critical quantum systems.  

To explicitly break the $SU(2)$, we supplement the Hamiltonian~\eqref{eq:ham} with an external {\it staggered} field $h$, which directly couples to the order parameter:
\begin{equation}\label{eq:ham-with-hstag}
{\cal H}_h =  {\cal H} - h\sum_{R} (-1)^{x+y}S^z_{R}\,,
\end{equation}
where $R=(x,y)$. Then, for any $h \ne 0$, the ground state has a finite correlation length and develops a finite staggered magnetization $m=m(h,\alpha)$ in response 
to the staggered field $h$. The response for $h \to 0^+$ is different within magnetically ordered or disordered phases. In both cases, the magnetization is an analytic 
function of $h$, with two distinct regimes:
\begin{eqnarray}
m(\alpha,h) &=& a(\alpha) h +O(h^2)  \qquad\mathrm{for}\;\alpha<\alpha_c,\, h\to0^+, \\
m(\alpha,h) &=& m(\alpha,0) +O(h)    \qquad\mathrm{for}\;\alpha>\alpha_c,\, h\to0^+,
\end{eqnarray}
where $a(\alpha)$ is a suitable constant. By contrast, at criticality (i.e., for $\alpha=\alpha_c$), the magnetization is not analytic and the response follows a 
power-law behavior:
\begin{equation}
m(\alpha_c,h) \propto h^{1/\delta}\;, \;\; \mathrm{for}\; h\to0^+\,,
\end{equation}
where $\delta$ is a critical exponent, which only depends on the universality class of the phase transition. The best estimate of $\delta$ (within the expected 
universality class of the classical three-dimensional $O(3)$ Heisenberg model) $1/\delta=0.20916$~\cite{chester2021}. 

To probe the system's response at fixed $\alpha$, we define the logarithmic derivative
\begin{equation}
[1/\delta](\alpha,h) = \frac{\partial \log m(\alpha,h)}{\partial \log h}\,,
\end{equation}
which is usually referred to as {\it running exponent}. For each value of $h$, we optimize the $U(1)$-symmetric iPEPS {\it Ansatz} and compute the average staggered 
magnetization $m(\alpha,h)$. Then, we estimate the logarithmic derivative by using finite differences. The QMC results are based on a direct improved estimator for 
the running exponent, see Ref.~\cite{demidio2021}. The outcomes are shown in Fig.~\ref{fig:hz-alpha}, where the running exponent and the correlation length $\xi_x$ 
of iPEPS are reported for the three different regimes (with $D=2,\ldots,6$).

In the gapped phase (for $\alpha=0.2$), the magnetization is linear in $h$ (linear response) and, therefore, the running exponent saturates at $1$ for $h \to 0$, 
as also obtained numerically from both QMC and iPEPS data (with $D \geq 4$). The iPEPS correlation length saturates for $h\lesssim 0.005$, with small finite-$D$ 
corrections. Indeed, in the gapped phase, even finite-$D$ iPEPS provide precise description of the system as the area law is upheld, which is supported by a direct 
comparison between iPEPS (with $D \geq 4$) and QMC (for $L \geq 32$). The $D=3$ iPEPS cannot faithfully capture gapped phase at $\alpha=0.2$, since it retains finite 
$m$ and thus responds more akin to the N\'eel phase.

Within the magnetically ordered phase (for $\alpha=0.4$), the magnetization saturates to a finite value and the running exponent goes to zero for $h \to 0$. Indeed, 
the logarithmic derivative $[1/\delta](\alpha,h)$ has a non-monotonic behavior, with a broad peak at intermediate values of the staggered field $h$. This feature is 
correctly captured by iPEPS. QMC calculations also reproduce the broad peak, even though at small values of $h$ a huge upturn is present due to size effects; 
therefore, for large enough system size $L$, the two methods converge to the same curve for sufficiently large values of the staggered field $h$. Here, it seems as 
though QMC simulations provide an upper bound of the thermodynamic limit, while iPEPS provide a lower bound of it. Still the presence of the broad peak is robust and 
appears in the region of $h$ where both $D \geq 5$ iPEPS and $L \geq 32$ QMC data are in excellent agreement. 

At the critical point $\alpha=\alpha_c$, the running exponent is expected to converge towards $1/\delta=0.20916$, as mentioned before. The available iPEPS and QMC
calculations (for $D \leq 6$ and $L \leq 128$, respectively) are still somewhat away from the saturation regime, however the numerical data monotonously increase, 
without the broad peak in the magnetically orderd phase. While the validity of the QMC results is limited by the finite size $L$, for iPEPS it is limited by the 
finite correlation length induced by the bond dimension $D$. Within iPEPS, as the field $h$ is decreased, the correlation lengths initially grow, but then flatten 
out (with large finite-$D$ corrections), in contrast to the expected diverging behavior of the gapless regime. From the direct comparison between QMC and iPEPS data, 
we can estimate that the iPEPS response for $D=6$ gives a faithful estimation of the exact result down to $h \approx 10^{-2}$. Then, for smaller fields, the running 
exponent data for iPEPS and large-$L$ QMC simulations start to deviate as iPEPS becomes increasingly biased by the induced finite correlation length.

\section{Conclusions and outlook}\label{sec:conclusions}

The analysis of quantum phase transitions by variational methods relies on a few aspects: first, a flexible {\it Ansatz} for the ground-state wave fuction, allowing
a description of different kinds of phases by tuning its parameters; second, a practical optimization scheme of such parameters to get the best approximate ground 
state; finally, a way to analyze the results, possibly extrapolating to the thermodynamic limit. Here, we have shown that symmetric tensor networks, when implemented 
with a variational optimization, allow for accurate description of the transition between a gapped paramagnet and gapless N\'eel antiferromagnet, thus overcoming the 
issues arising within the imaginary-time evolution approach that has been emphasized in a recent work~\cite{hasik2019}. The selection of the correct symmetry structure 
for tensors by the choice of their charges is crucial, as they determine the physical properties of the wave function. Most importantly, once the tensors have been 
variationally optimized, our simulations show that the physical observables, such as the energy or the order parameter, are robust to small variations of these charges. 
Although the region close to the critical point remains a challenge, it is in principle tractable by increasing the bond dimension $D$. Already with data up to $D=6$ 
and FCSL scaling, we could locate the critical point with an accuracy of about $5\%$, demonstrating the applicability of this analysis even in the case of two 
different length scales $\xi_x$ and $\xi_y$. Finally, inspired by the recent works of Refs.~\cite{rader2018,demidio2021}, we have included an external staggered field 
$h$ in the Hamiltonian and compared the iPEPS results with the ones obtained by QMC. Away from the critical point $\alpha>\alpha_c$ (i.e., within the N\'eel phase) 
the running exponent $[1/\delta](\alpha,h)$ shows a broad peak at finite field $h$, indicative of magnetic order~\cite{demidio2021}. The robustness of the peak can 
be established by locating the inflection point in the growing iPEPS correlation length as the external field $h$ is decreased. Peaks at fields larger than inflection 
point represent genuine features of the system, while finite-$D$ effects may generate spurious peaks at fields smaller than the inflection point. Our findings 
corroborate the fact that the analysis of the running exponent is a useful diagnosis within the antiferromagnetic regime, away from the critical point. This proof of 
concept analysis shows the potential of the method put forward in Ref.~\cite{demidio2021} for future application of iPEPS, e.g., on questions of the stability of 
quantum spin liquids.

\begin{acknowledgments}
J.H. thanks D. Poilblanc and P. Corboz for valuable discussions. This project has received funding from the European Research Council (ERC) under the European Union's 
Horizon 2020 research and innovation programme (grant agreement No 101001604), TNSTRONG ANR-16-CE30-0025, TNTOP ANR-18-CE30-0026-01 grants awarded from the French 
Research Council, the Austrian Science Fund FWF project I-4548, and the SFB BeyondC Project No. F7108-N38. This work was also granted access to the HPC resources of CALMIP supercomputing center under the 
allocations 2017-P1231 and 2021-P0677.
\end{acknowledgments}

\bibliography{bibliography}

\clearpage
\appendix

\section{Comparison among $U(1)$ classes}\label{sec:app-u1classes}

First of all, we discuss the preparation of the initial $U(1)$-symmetric states for the optimization by imaginary-time evolution. We consider two scenarios of evolution 
by starting from simple $U(1)$-symmetric product states: a classical N\'eel antiferromagnet ({\it NEEL}) and a product state of singlets placed on the rungs of ladders 
({\it VBS}). The imaginary-time evolutions were performed by using two-site simple update (SU) with the second-order Trotter approximation. To track the convergence of 
SU, we evaluate energy using corner-transfer matrix environments, with modest environment dimension $\chi \approx D^2$. Starting with the time step $\delta \tau = 0.02$,
the state is evolved until an energy increase is observed, in which case the SU step is not performed and the time step is halved $\delta \tau \rightarrow \delta\tau/2$.
The evolution ends once the time step becomes smaller than $\delta \tau=10^{-8}$.

\begin{figure}[tbp]
\includegraphics[width=\columnwidth]{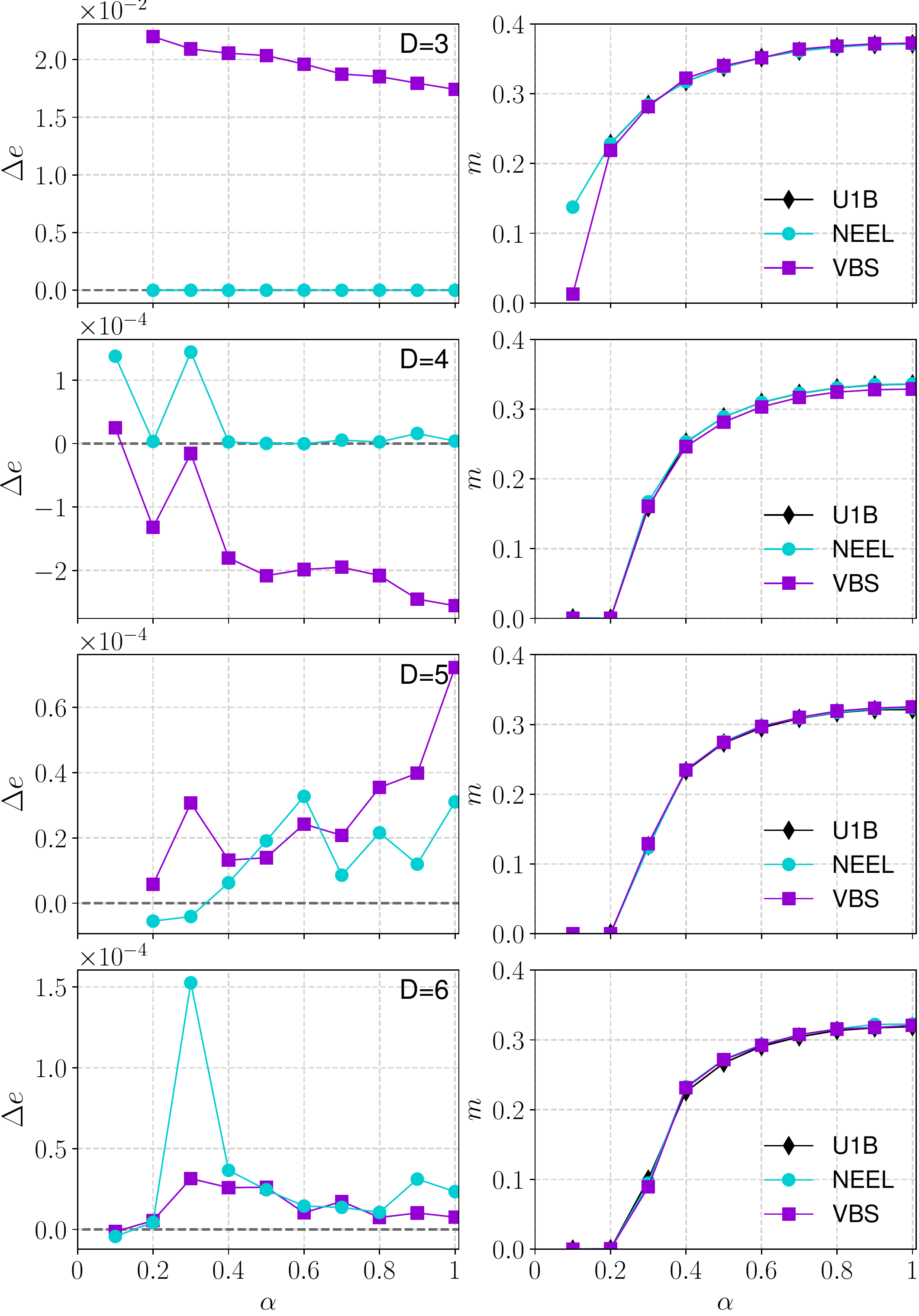}
\caption{\label{fig:u1-classes-opt-AD} 
Left panels: energy per site obtained optimizing {\it NEEL} and {\it VBS} states with respect to the one obtained within the {\it U1B} class, i.e., 
$\delta e=|e({\it NEEL})-e({\it U1B})|$ (circles) and $\delta e=|e({\it VBS})-e({\it U1B})|$ (squares) for different values of the bond dimension $D$. Right panels:
magnetization $m$ obtained from {\it NEEL}, {\it VBS}, and {\it U1B} states. Observables are evaluated at finite $\chi$, i.e., $\chi=71$ for $D=3$, $\chi=64$ for 
$D=4$, $\chi=50$ for $D=5$ and $\chi=72$.}
\end{figure}

The $U(1)$-structure of the evolved {\it NEEL} and {\it VBS} states show slight variations in charges $\vec{v}^\gamma_j$. Moreover, depending on bond dimension $D$ 
and $\alpha$, the charges do not necessarily respect spatial symmetries of the model~\eqref{eq:ham}. For example, at $D=5$ and $6$ the charges on left ($j=l$) and 
right ($j=r$) bonds of tensors come out different. Then, we perform a further optimization of the {\it NEEL} and {\it VBS} states by using gradient descent and compare 
the resulting physical observables with the ones obtained from optimal states with identical charges $\vec{v}^\gamma_j$ for all bonds $j$), here denoted as {\it U1B} 
class. The results for the energy per site $e$ and the local magnetization $m$ are shown in Fig.~\ref{fig:u1-classes-opt-AD}. The data for $D=3$ show that optimized 
{\it VBS} states have substantially worse energies than {\it NEEL} and {\it U1B} states. Instead, for $D>3$, all these states have very similar energies, differences 
being at most of order $2 \times 10^{-4}$. The order parameter displays a similar behavior. A minor exception is represented by the $D=4$ VBS case, which displays 
slight breaking of translation symmetry along $y$-axis even in the $\alpha=1$ limit. In particular, the nearest-neighbor spin-spin correlations show staggered pattern 
with $\langle\vec{S}_{x,y}\cdot\vec{S}_{x,y+1}\rangle - \langle\vec{S}_{x,y+1}\cdot\vec{S}_{x,y+2}\rangle \approx 3\times10^{-4}$. Overall, the SU-evolved states 
provide reasonable accurate initial states for the variational optimization of $U(1)$-symmetric iPEPS. Irrespective of the differences in the $U(1)$ charges, the final 
variational minima for different classes give quantitatively the similar physical picture. 

\begin{figure}[tbp]
\includegraphics[width=\columnwidth]{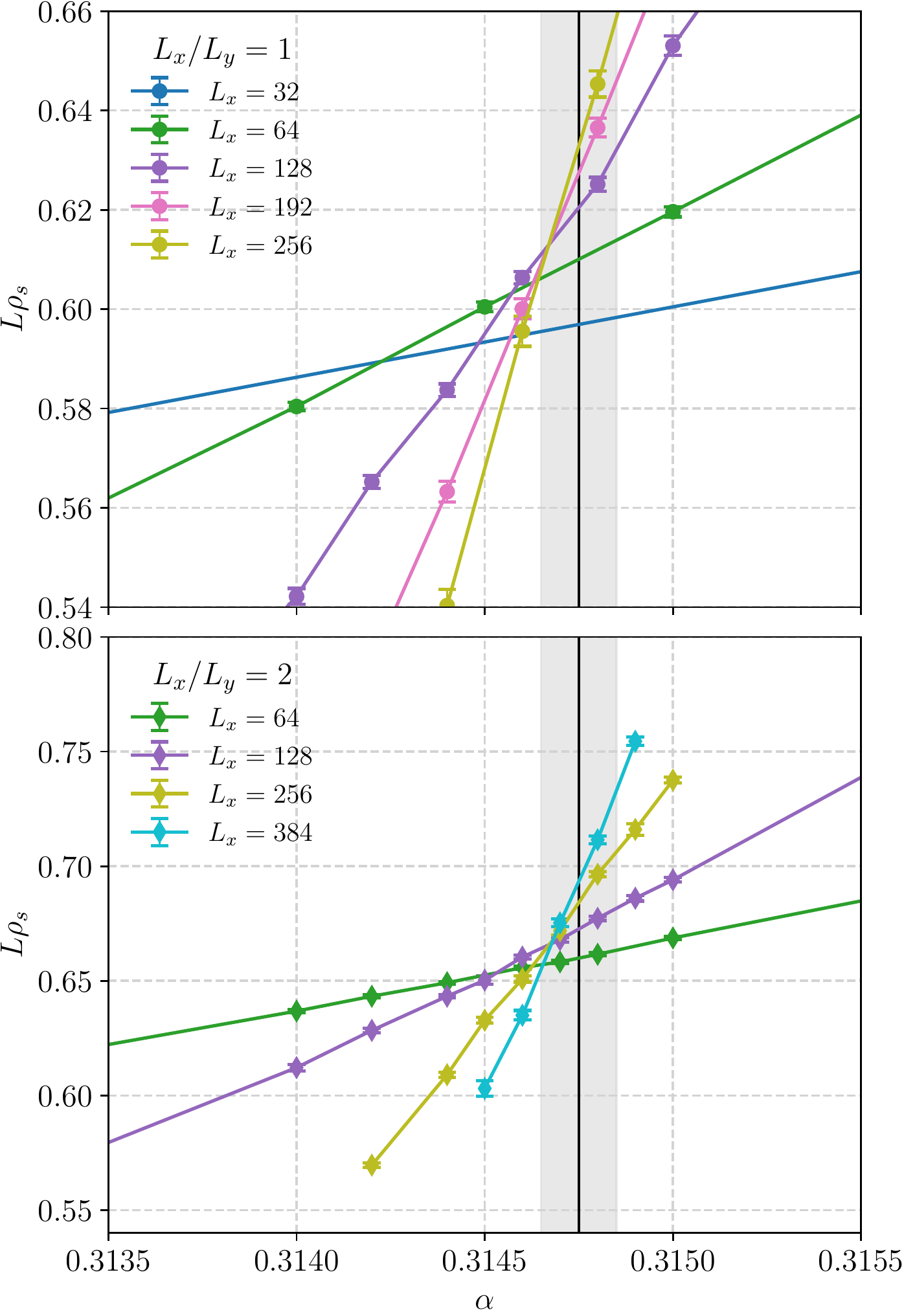}
\caption{\label{fig:qmc-alphac}
QMC results for the scaling analysis of the spin stiffness $\rho_s$ as a function of $\alpha$ for system with $L_x=L_y$ (upper panel) and $L_x=2L_y$ (lower panel). 
The temperature has been chosen as $T=1/(2L_x)$ to ensure convergence to the ground-state properties. The vertical black line and shaded area indicate 
$\alpha_c = 0.31475(10)$ estimate and its uncertainty.}
\end{figure}

\begin{figure}[bp]
\centerline{\includegraphics[width=0.9\columnwidth]{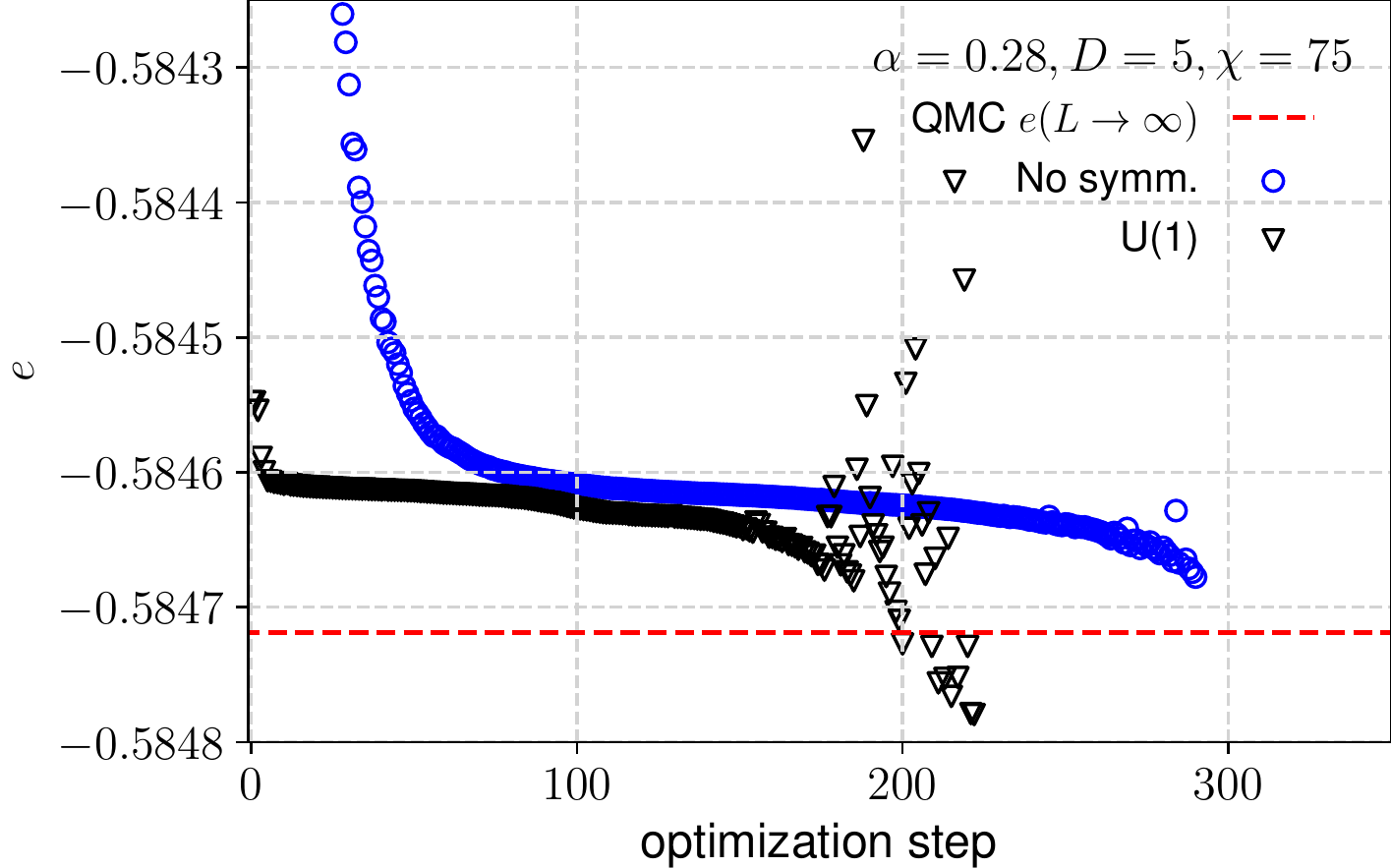}}

\vspace{5mm}
\centerline{\includegraphics[width=0.9\columnwidth]{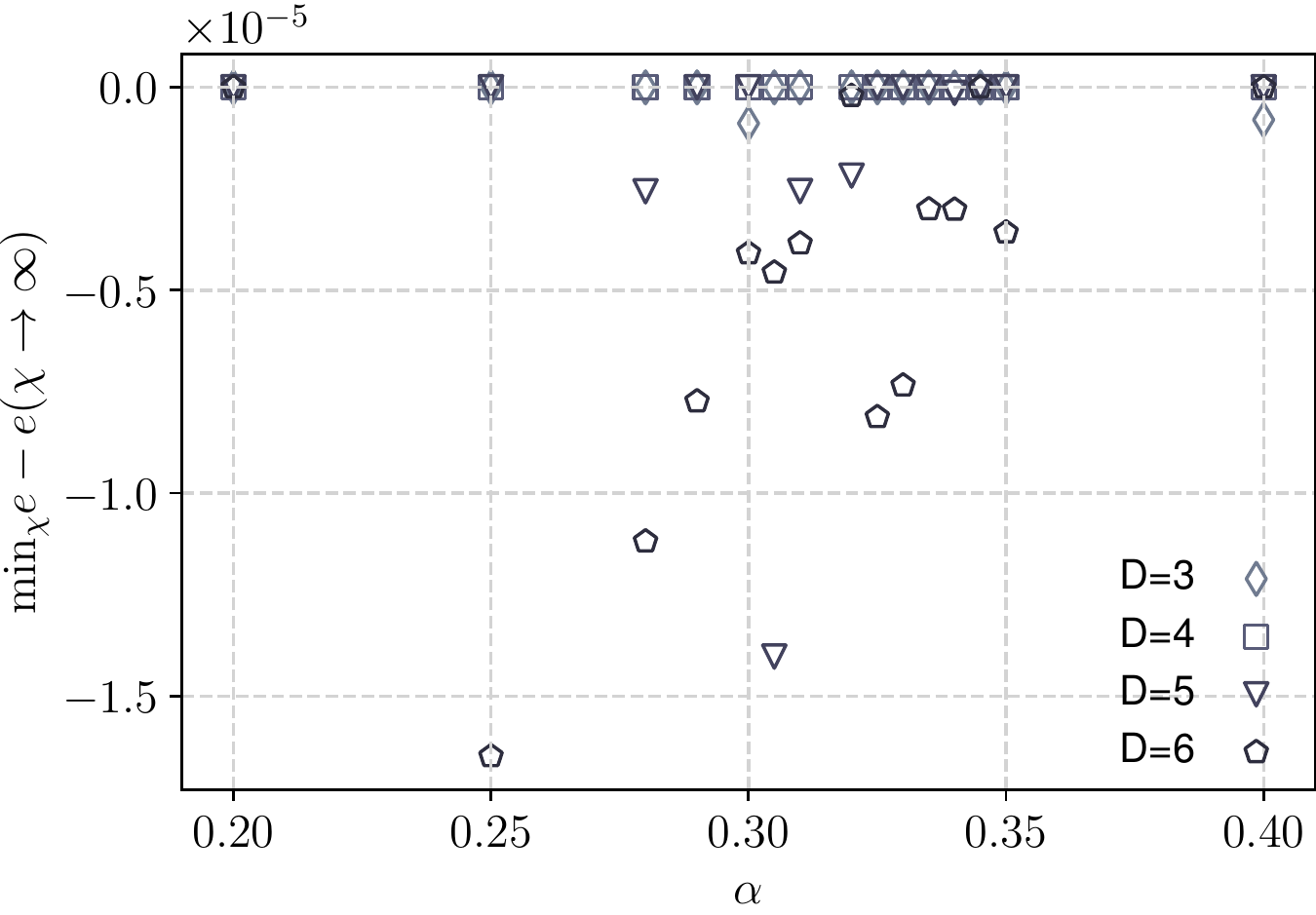}}
\caption{\label{fig:optimization-instability}
Upper panel: example of optimizations without line search at $\alpha=0.28$ (with $D=5$ and $\chi=75$) both with and without explicit $U(1)$ symmetry. Optimizations 
are initialized from the $\alpha=0.3$. The red line shows the comparison with QMC extrapolation. Lower panel: difference between thermodynamic estimate of the energy 
($\chi \to \infty$ limit) and its minimal value for fixed environment dimension $\chi$ for optimized $U(1)$ symmetric states. The optimizations were done for $D=3$, 
$4$, $5$, and $6$ at $\chi_{opt}=54$, $64$, $100$, and $108$, respectively. In all cases, the line search was employed.}
\end{figure}

\section{Size scaling analysis of the transition by the QMC technique}\label{sec:app-qmc-alphac} 

Here, we provide some standard details on  locating the transition point $\alpha_c$ from QMC simulations. To estimate $\alpha_c$, we have computed the spin stiffness 
$\rho_s$, defined using the winding number: 
\begin{equation}
    \rho_s = \frac{T}{2}\langle \sum_i (W_i^x)^2 + (W_i^y)^2 \rangle
\end{equation}
where $i$ runs over loops in the space-time SSE configurations and the brackets denote the Monte Carlo average.

The results for different system sizes $L_x \times L_y$ are shown in Fig~\ref{fig:qmc-alphac}. At criticality, the scaled spin stiffness $L_x\rho_s$ is constant 
(taking the dynamical exponent $z=1$ at the transition), thus allowing to estimate the location of the transition at $\alpha_c=0.31467(1)$. Note that the Binder 
cumulant crossing also leads to a similar critical value (data not shown). In addition, we have also checked that data collapse of $\rho_s$ and magnetization are 
compatible with known exponents $\nu=0.707$ and $\beta=0.3689$.

\section{Instability in optimization}\label{sec:opt-instability}

The optimization of the iPEPS wave function is performed by using the steepest-descent algorithm, implemented with the line search method (i.e., by using a step 
length that is adapted along the optimization procedure)~\cite{numerical}. In the proximity of the critical point, performing the line search is crucial to obtain 
stable  variational optimizations, especially for the cases with $D=5$ and $6$. In Fig.~\ref{fig:optimization-instability} (upper panel), we show typical cases done
without line search, such that unstable regimes, with an erratic behavior of the energy, are obtained. The same kind of problem appears in simulations with or 
without the $U(1)$ symmetry.

When performing the optimization, we always work at constant environment bond dimension $\chi_{opt}$. The states obtained in the erratic regime can show energies 
$e(\chi_{opt})$, which are lower than the reference QMC results. This is possible because the iPEPS thermodynamic estimate of energy, for which the variational 
principle holds, is obtained only in the limit of $\chi \to \infty$. When the line search is employed, it suppresses the erratic regime as the energy is not allowed 
to increase in course of optimization.  In this case, we asses the severity of this instability by comparing the lowest energies that are realized in the vicinity 
of $\chi_{opt}$ with the energy obtained from the $\chi\rightarrow\infty$ limit. The results for $D=3,\ldots,6$ and $0.2 \le \alpha \le 0.4$ are shown in 
Fig.~\ref{fig:optimization-instability} (lower panel). This behavior resembles the overtraining, often encountered in the optimization of artificial neural networks, 
since the variational optimization of iPEPS does not directly optimize the thermodynamic estimate of energy $e(\chi\rightarrow\infty)$, but only its finite-$\chi$ 
approximation. In the previous study of the $J_1-J_2$ model (using highly-constrained single-site iPEPS {\it Anstaz}~\cite{hasik2021}), $e(\chi)$ was a monotonically 
decreasing function already from modest values of $\chi$. Here, we instead observe that $e(\chi)$ attains shallow spurious minima, with depth at most $O(10^{-5})$, 
compared to corresponding $e(\chi\rightarrow\infty)$ thermodynamic estimates.

\end{document}